# A topological transition-induced giant transverse thermoelectric effect in polycrystalline Dirac semimetal Mg$_3$Bi$_2$

Tao Feng[1]†, Panshuo Wang[2]†, Zhijia Han[1], Liang Zhou[2], Zhiran Wang[1], Wenqing Zhang[1]*, Qihang Liu[2]*, Weishu Liu[1,3]*

[1] Department of Materials Science and Engineering, Southern University of Science and Technology, Shenzhen 518055, China

[2] Department of Physics and Shenzhen Institute for Quantum Science & Engineering, Southern University of Science and Technology, Shenzhen 518055, China

[3] Guangdong Provincial Key Laboratory of Functional Oxide Materials and Devices, Southern University of Science and Technology, Shenzhen 518055, China

*Corresponding author. Email: liuws@sustech.edu.cn, liuqh@sustech.edu.cn, zhangwq@sustech.edu.cn

**Abstract**

To achieve thermoelectric energy conversion, a large transverse thermoelectric effect in topological materials is crucial. However, the general relationship between topological electronic structures and transverse thermoelectric effect remains unclear, restricting the rational design of novel transverse thermoelectric materials. Herein, we demonstrate a topological transition-induced giant transverse thermoelectric effect in polycrystalline Mn-doped $Mg_{3+\delta}Bi_2$ material, which has a competitively large transverse thermopower (617 μV/K), power factor (20393 $μWm^{-1}K^{-2}$), magnetoresistance (16600%), and electronic mobility (35280 $cm^2V^{-1}s^{-1}$). The high performance is triggered by the modulation of chemical pressure and disorder effects in the presence of Mn doping, which induces the transition from a topological insulator to a Dirac semimetal. The high-performance polycrystalline Mn-doped $Mg_{3+\delta}Bi_2$ described in this work robustly boosts transverse thermoelectric effect through topological phase transition, paving a new avenue for the material design of transverse thermoelectricity.

**Keywords**

Transverse thermoelectric, Topological phase transition, Magnetoresistance, Chemical pressure, Disorder effect, Polycrystalline $Mg_{3+\delta}Bi_2$:$Mn_x$

## 1. Introduction

Transverse thermoelectricity based on the Nernst effect has proven to be compelling potential for application in energy harvesting technology [1-3]. The voltage drop is generated along the direction perpendicular to the temperature gradient as a magnetic field is applied in the normal direction of the vector plane of the temperature gradient and voltage drop [4]. Such scenario is not only conducive to decoupling electron and phonon transport, but also helpful in simplifying thermoelectric devices [5,6]. At present, most of the reported transverse thermoelectric materials belong to the category of topological semimetal with high carrier mobility, which is very different from the classic thermoelectric semiconductor, which is confined to a narrow band gap [7-14]. For example, the topological Dirac semimetal $ZrTe_5$ was shown to have a high electron mobility (< 64000 $cm^2V^{-1}s^{-1}$), leading to an evident transverse thermoelectric effect with a robust transverse power factor (10314 $μWm^{-1}K^{-2}$ at 115 K) [7]. Similarly, the Dirac semimetal $Cd_3As_2$ is known to have a large transverse thermopower (115 μV/K) and power factor (5000 μWm$^{-1}$K$^{-2}$ at 350 K), largely due to the charge compensation of pertinent electrons and holes and their high mobilities [8,9]. A third alternative, the Weyl semimetal NbP has also exhibited strong transverse thermoelectric responses (800 μV/K) and ultra-high carrier mobility ($5 \times 10^6$ cm$^2$V$^{-1}$s$^{-1}$) [10,14]. However, although efforts has been made on the transverse thermoelectric properties of topological semimetal materials, the general relationship between topological electronic structures and transverse thermoelectric properties remains unclear; thus, a simple and effective way to manipulate the performance of transverse thermoelectric materials remains elusive.

In recent years, $Mg_{3+\delta}Bi_2$ has drawn intense attention as a promising thermoelectric material, exhibiting excellent performance in traditional thermoelectricity [15]. It has relatively low transverse thermoelectric performance, however, due to its low carrier mobility (~450 cm$^2$V$^{-1}$s$^{-1}$) [16,17], which has been attributed to the formation of Mg vacancy defects during the synthesis process [18]. Previously, we demonstrated that the defect engineering method (i.e. Fermi energy tuning) was effective in suppressing Mg vacancies and optimizing carrier mobility [19]. The electron mobility increased by an order of magnitude, boosting magnetoresistance (MR) by 940% and vastly improving both the transverse thermopower (127 μVK$^{-1}$) and transverse power factor (2182 μWm$^{-1}$K$^{-2}$) at 13.5 K and 6 Tesla, rendering the $Mg_{3+\delta}Bi_2$ a promising transverse thermoelectric material. However, two open questions remained. Is it possible to further boost the transverse thermoelectric effect by regulating the topological electronic structure in topological materials? If so, is there a simple way to search new promising candidates?

Owing to the linear-dispersed bands of Dirac semimetals, their massless low-energy excitations can dramatically enhance carrier mobility. Thus, a topological transition of $Mg_{3+\delta}Bi_2$ from a $Z_2$ topological insulator to a Dirac semimetal phase would be desirable. In this work, we report that the as-fabricated $Mg_{3+\delta}Bi_2$:Mn$_x$ has a large transverse thermopower (617 μV/K at 14 K and 14 Tesla) and power factor (20393 μWm$^{-1}$K$^{-2}$ at 14 K and 14 Tesla), the highest value among the reported polycrystalline transverse thermoelectric materials. The vastly improved electronic mobility (35280 cm$^2$V$^{-1}$s$^{-1}$ at 2 K) and MR (16600% at 2 K and 14 Tesla) outperformed the defect-engineering-optimized $Mg_{3+\delta}Bi_2$ by 7 and 17 times, respectively. Through first-principles calculations of pristine and doped $Mg_{3+\delta}Bi_2$, we verified the topological phase transition from $Z_2$ topological insulator to Dirac semimetal via Mn doping as a result of the combined effects of the negative chemical pressure and disorder effect (Fig. 1A). As anticipated, our MR measurements and transport calculations also showed a clear positive relation between transverse thermopower and MR. Our work provides a technical route for exponentially boosting transverse thermoelectricity through the direct engineering of topological electronic structures and identifies high non-saturated MR as a critical feature of promising candidates for future high-performance transverse thermoelectric materials.

## 2. Results and Discussion

### 2.1. Giant transverse thermoelectric effect

The as-fabricated polycrystalline Mn-doped material is denoted as $Mg_{3+\delta}Bi_2$:$Mn_x$ (x = 0, 0.025, 0.075, 0.1, 0.125, 0.15; δ indicates Mg-rich condition), which was synthesized by combined mechanical alloying and spark plasma sintering [19]. The nominal compositions of these samples are presented in Table S1. The Rietveld refinements of the as-fabricated $Mg_{3+\delta}Bi_2$:$Mn_x$ samples with a space group of $P\bar{3}m1$ showed that both the *a* axis and *c* axis increased with Mn content (Fig. 1B), and the corresponding cell volume *V* gradually increased from 137.8 Å$^3$ to 142.8 Å$^3$ as the Mn content increased from x = 0 to x = 0.15. The Mn dopant expanded the lattice, inducing negative chemical pressure in the $Mg_{3+\delta}Bi_2$ matrix. We believed that the negative chemical pressure could be similar to tensile strain, inducing a topological transition from a $Z_2$ topological insulator into a Dirac semimetal. The temperature-dependent transverse thermopower of the as-fabricated $Mg_{3+\delta}Bi_2$:$Mn_x$ (x = 0.1) polycrystalline bulk under different magnetic fields (H) exhibits a peak near 14 K (Fig. 1C). The peak value increases up to 617 μV/K at 14 Tesla, which is 5 times higher than that reported for polycrystalline $Mg_{3+\delta}Bi_2$ [19] and also much higher than the peak values reported for other polycrystalline transverse thermoelectric materials (Fig. 1D) (e.g. $Re_4Si_7$, $Mg_2Pb$, etc.) [8,11,20-24].

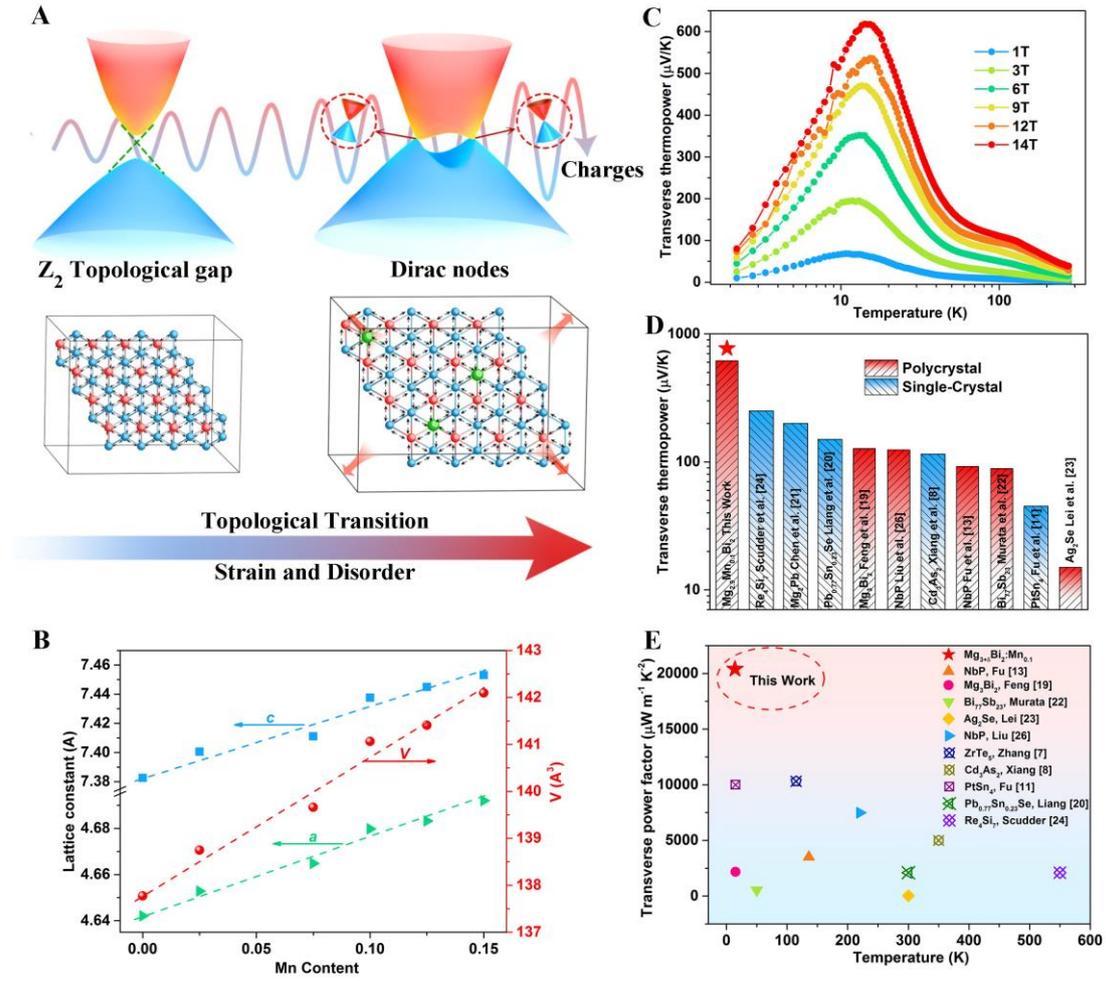

**Fig. 1. Transverse thermoelectric performance of polycrystalline $Mg_{3+\delta}Bi_2:Mn_x$ (x = 0.1).** (**A**) A schematic of the topological phase transition of $Mg_{3+\delta}Bi_2:Mn_x$ from a $Z_2$ topological insulator to a Dirac semimetal, resulting from the combined effects of negative chemical pressure and disorder. (**B**) The *a* axis, *c* axis and cell volume of the obtained Mn-doped samples according to the Rietveld refinements with a space group of $P\bar{3}m1$. (**C**) Transverse thermopower under different magnetic fields in a temperature range of 2-275 K. (**D**) A transverse thermopower comparison diagram of promising polycrystalline (red) and single-crystal (blue) materials. (**E**) A transverse power factor comparison diagram among promising transverse thermoelectric materials, where the filled and unfilled symbols represent polycrystalline and single-crystal materials, respectively.

Fig. 1E further compares the transverse power factor (maximum output power density in practical applications [25]) of the as-fabricated $Mg_{3+\delta}Bi_2:Mn_x$ (*x* = 0.1) polycrystalline bulk with reported transverse thermoelectric materials. For all transverse thermoelectric materials, the transverse power factor was defined as $PF_T = S_T^2\sigma$, where $S_T$ is the transverse thermopower perpendicular to the temperature gradient and $\sigma$ is the regular electrical conductivity. Since $S_T$ and $\sigma$ usually have an oppositive correction to the magnetic field strength, the $PF_T$ showed a magnetic-field-dependent peak (Fig. S3). In our investigated range (0-14 Tesla), the as-fabricated $Mg_{3+\delta}Bi_2$:$Mn_{0.1}$ showed a maximum value of 20393 $\mu Wm^{-1}K^{-2}$ under a 6 Tesla magnetic field, which was almost an order of magnitude higher than the 2182 $\mu Wm^{-1}K^{-2}$ value reported for the $Mg_{3+\delta}Bi_2$ sample [19]. This value was also much higher than that of all polycrystalline materials and most single-crystal materials reported to date, including the Dirac semimetals $ZrTe_5$ [7], $PtSn_4$ [11], $Pb_{0.77}Sb_{0.23}Se$ [20], $Cd_3As_2$ [8], and NbP [10,13,14,26]; the semimetals $Mg_2Pb$ [21] and $Bi_{77}Sb_{23}$ [22]; and the narrow-band semiconductors $Ag_{2(1+x)}Se$ [23] and $Re_4Si_7$ [24]. We also note that this value is slightly lower than that for the single-crystal Weyl semimetal $WTe_2$ [27]. The polycrystalline $Mg_{3+\delta}Bi_2$:$Mn_{0.1}$, however, had obvious advantages in the preparation of large-size bulk and easy regulation of Fermi level, thus showing a promising realistic application.

## 2.2. An inverse relationship between $S_T$ and $\sigma$ under a magnetic field

Classic thermoelectric materials are characterized by a carrier concentration-dependent inversion between $S$ and $\sigma$ [28], and transverse thermoelectric materials are similarly characterized by the magnetic field strength, which also connects the $S_T$ and $\sigma$. As the charge carriers diffuse along the temperature gradient, the Lorentz force bends the trace of the charge flux due to the magnetic field—that is, the stronger the magnetic field, the larger the transverse charge flux, resulting in a higher transverse thermopower. For the $\sigma$, a stronger magnetic field leads to a less charge carrier transport along the longitudinal direction, corresponding to decreased $J_L$. (Fig. 2A). To verify the inverse relationship between $S_T$ and $\sigma$, we measured the MR of the as-fabricated $Mg_{3+\delta}Bi_2$:$Mn_x$ under different magnetic fields. Fig. 2B shows a magnetic-field-dependent MR of the $Mg_{3+\delta}Bi_2$:$Mn_{0.1}$ sample at different temperatures (2-300 K). Unsurprisingly, the MR at each temperature follows a linear unsaturated trend up to 14 Tesla, which was consistent with the previous reported $Mg_{3+\delta}Bi_2$ sample [19] and the Dirac semimetals $Pb_{1-x}Sn_xSe$ and $Cd_3As_2$ [8,22]. The corresponding MR of other as-fabricated $Mg_{3+\delta}Bi_2$:$Mn_x$ are shown in Figs. S2A-S2E. Fig. 2C plots the MR of $Mg_{3+\delta}Bi_2$:$Mn_x$ at 2 K and 14 Tesla as a function of the Mn content, showing a dramatic increase in MR from 940% ($x = 0$) to a peak at 16600% ($x = 0.1$), followed by a slight decrease to 10300% ($x = 0.15$). These significant changes suggest that the chemical pressure caused by the Mn dopant alters the topology of the electronic structure.

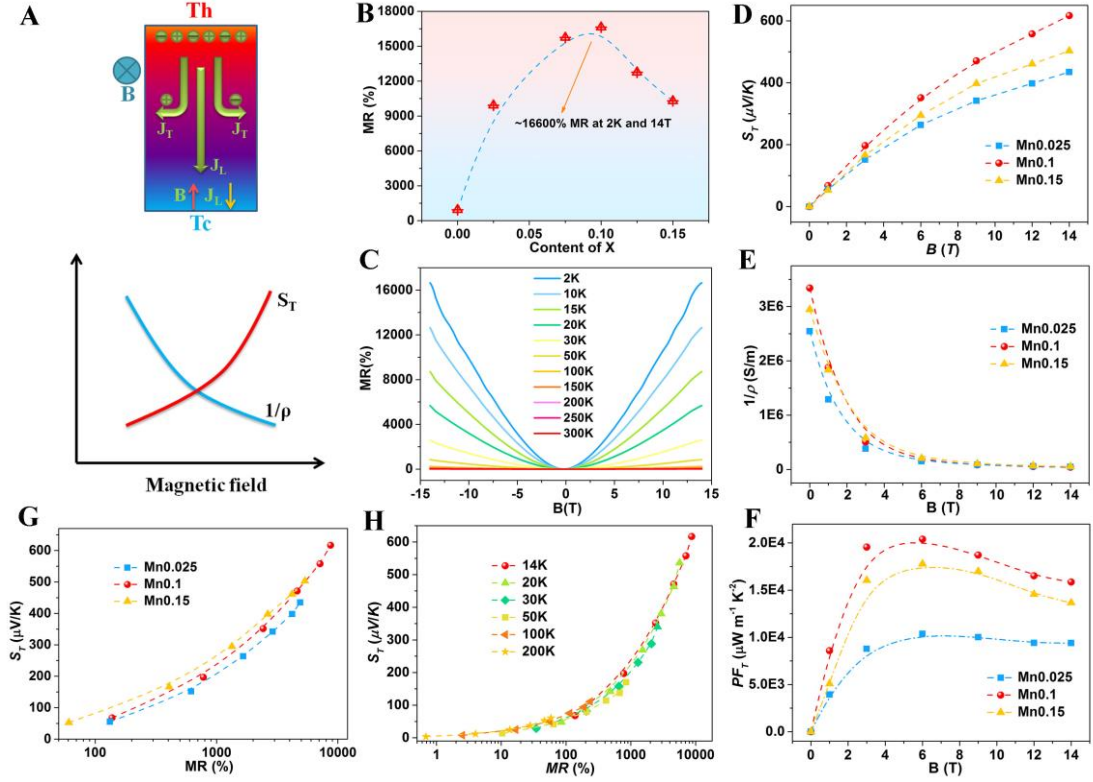

**Fig. 2. Inverse relationship between $S_T$ and $\sigma$.** (**A**) A diagram of transverse thermoelectricity and the inverted relationship between $S_T$ and $1/\rho$. (**B**) The MR of $Mg_{3+\delta}Bi_2$:$Mn_{0.1}$ at different temperatures in a magnetic field between −14 and 14 Tesla. (**C**) The Mn content-dependent MR of as-fabricated polycrystalline $Mg_{3+\delta}Bi_2$:$Mn_x$ ($0 \leq x \leq 0.15$) at 2 K and 14 Tesla. The magnetic field-dependent (**D**) $S_T$, (**E**) $1/\rho$, and (**F**) $PF_T$ for $Mg_{3+\delta}Bi_2$:$Mn_x$ ($x = 0.025, 0.1, 0.15$). The relationship between $S_T$ and MR for (**G**) $Mg_{3+\delta}Bi_2$:$Mn_x$ (x = 0.025, 0.1, 0.15) samples at 14 K and for (**H**) $Mg_{3+\delta}Bi_2$:$Mn_{0.1}$ under different temperatures.

Fig. 2D-2F shows the magnetic-field-dependent transverse thermoelectric properties of the as-fabricated $Mg_{3+\delta}Bi_2$:$Mn_x$ (x = 0.025, 0.1, 0.15), which demonstrated similar magnetic-field-dependent behavior. As the magnetic-field increased from 0 to 14 Tesla, the $S_T$ increased from 0 μV/K for all $Mg_{3+\delta}Bi_2$:$Mn_x$ to 435 μV/K ($Mg_{3+\delta}Bi_2$:$Mn_{0.025}$), 617 μV/K ($Mg_{3+\delta}Bi_2$:$Mn_{0.1}$), and 503 μV/K ($Mg_{3+\delta}Bi_2$:$Mn_{0.15}$), respectively. The $Mg_{3+\delta}Bi_2$:$Mn_{0.1}$ sample exhibited the highest transverse thermopower in the whole magnetic field range. Unlike the $S_T$, the $1/\rho$ gradually decreased with the magnetic field for $Mg_{3+\delta}Bi_2$:$Mn_x$ (x = 0.025, 0.1, 0.15) samples

(Fig. 2E), which we attribute to the deflection of carriers by Lorentz force. Using the results of the magnetic-field-dependent $S_T$ and $1/\rho$, we calculated the $PF_T$. Fig. 2F plots the magnetic-field-dependent $PF_T$ for all three samples, which initially increased and then decreased, reaching the highest $PF_T$ values at 6 Tesla: 10370 μWm$^{-1}$K$^{-2}$ (Mg$_{3+\delta}$Bi$_2$:Mn$_{0.025}$), 20393 μWm$^{-1}$K$^{-2}$ (Mg$_{3+\delta}$Bi$_2$:Mn$_{0.1}$), and 17776 μWm$^{-1}$K$^{-2}$ for Mg$_{3+\delta}$Bi$_2$:Mn$_{0.15}$.

Fig. 2G shows a clear positive correlation between $S_T$ and MR for Mg$_{3+\delta}$Bi$_2$:Mn$_x$ samples with different Mn content at 14 K. For Mg$_{3+\delta}$Bi$_2$:Mn$_{0.025}$, the $S_T$ increased from 55 μV/K (132% MR) to 435 μV/K (4900% MR); for Mg$_{3+\delta}$Bi$_2$:Mn$_{0.1}$, the $S_T$ increased from 68 μV/K (138% MR) to 617 μV/K (8710% MR); and for Mg$_{3+\delta}$Bi$_2$:Mn$_{0.15}$, the $S_T$ increased from 52 μV/K (60% MR) to 503 μV/K (5350% MR). The MR-dependent $S_T$ under different temperatures for the Mg$_{3+\delta}$Bi$_2$:Mn$_{0.1}$ sample (Fig. 2H) shows a similar correlation, which further solidifies the inverse relationship existed between $S_T$ and $1/\rho$ under magnetic field.

**2.3. Topological phase transition induced by Mn doping**

We next performed first-principles calculations [19,29-32] to illustrate that the significantly enhanced transverse thermoelectric performance originated from the negative chemical pressure and disorder effect upon Mn doping. Fig. 3A and 3B show the topological phase transition of pristine Mg$_3$Bi$_2$ under 1% tensile strain. Without spin-orbit coupling (SOC), Mg$_3$Bi$_2$ is a type-II nodal-line semimetal with a nodal ring in the $k_x - k_y$ plane, whereas SOC opens a small gap (i.e., about 30 meV) around the nodal line (see Fig. 3A) [19,33-35]. The band component projections of the Mg-s orbitals (red dots) and the Bi-p orbitals (blue dots) characterized a clear band inversion between the $\Gamma_8^+$ bands and the $\Gamma_9^-$ bands at the $\Gamma$ point [36,37], where "+" and "−" indicate the parity of the Bloch wavefunctions. Based on the calculation of the evolution of the Wannier charge center for the six time-reversal invariant $\vec{k}$ planes [38] (see Fig. S7), the bulk Z$_2$ topological number was characterized as [1,000]. Thus, a strong topological insulator phase of Mg$_3$Bi$_2$ was demonstrated.

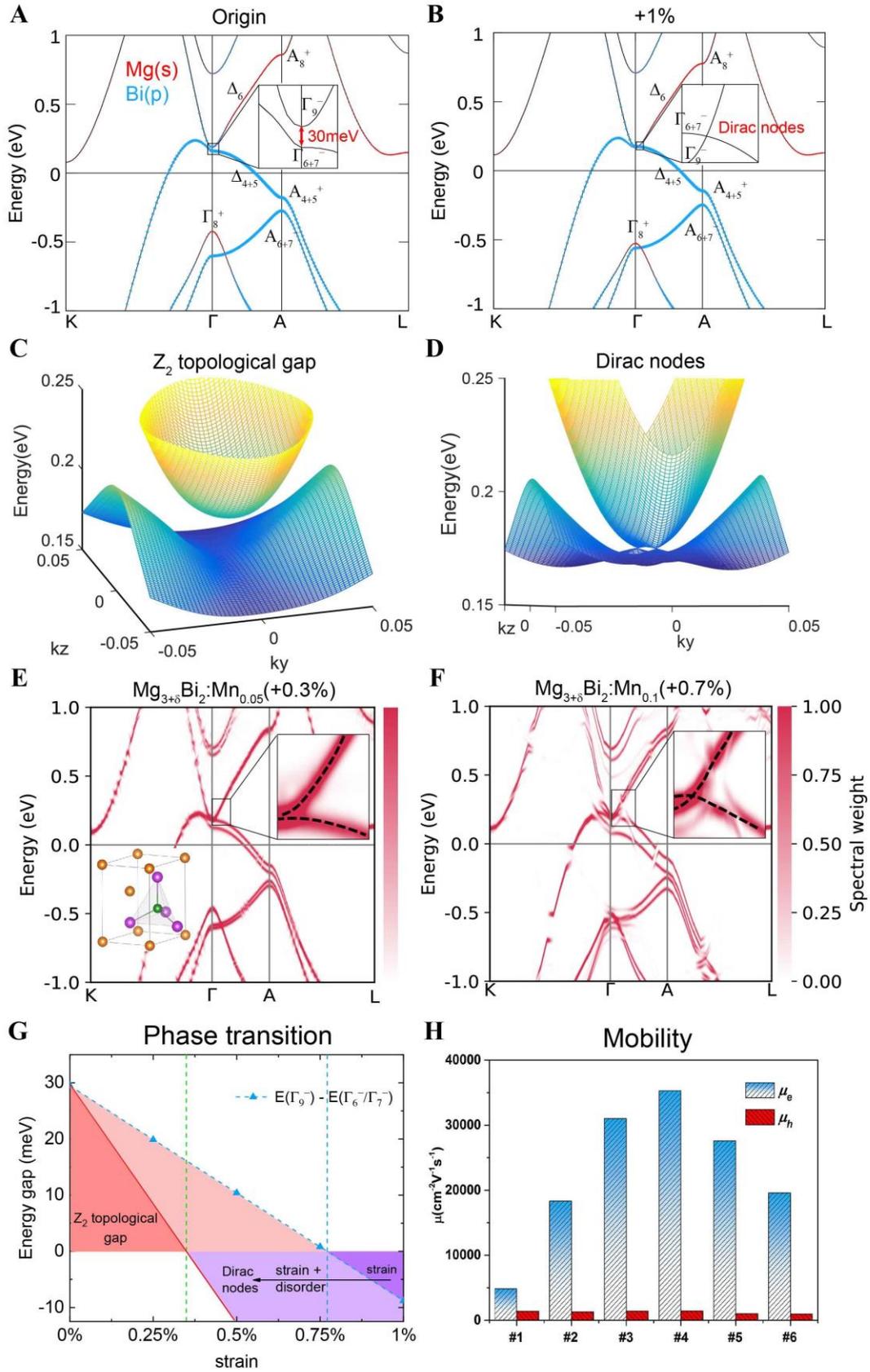

**Fig. 3. Topological phase transition of $Mg_{3+\delta}Bi_2:Mn_x$.** Band structures of pristine $Mg_3Bi_2$ with SOC for (**A**) original and (**B**) 1% tensile strain structures, with Mg-s components (red dots), Bi-p

components (blue dots), and irreducible representations and parities are also shown. Band evolution of the $k_x = 0$ plane from (**C**) original to (**D**) 1% tensile strain. Effective band structures of (**E**) $Mg_{3+\delta}Bi_2:Mn_{0.05}$ and (**F**) $Mg_{3+\delta}Bi_2:Mn_{0.1}$ with a tensile strain of 0.3% and 0.7%, respectively. (**G**) The energy differences between the $\Gamma_9^-$ and $\Gamma_{6+7}^-$ bands of $Mg_3Bi_2$, showing an increase in the tensile strain (blue dashed line), and the combined effects of the negative chemical pressure and disorder effect in $Mg_{3+\delta}Bi_2:Mn_x$ (red line), which accelerates the topological phase transition. (**H**) Electron and hole mobilities for $Mg_{3+\delta}Bi_2:Mn_x$ with various Mn contents.

According to the X-ray diffraction measurements, the lattice constants enlarged with the Mn doping (Fig. 1B), which is consistent with the fact that the ionic radius of $Mn^{2+}$ is larger than that of $Mg^{2+}$. We then investigated the electronic structure of pure $Mg_3Bi_2$ under homogeneous tensile strains (Fig. 3A-3D, 3G) to simulate the negative chemical pressure. With an increase in the tensile strain from 0% to 0.25%, 0.50%, and 0.75%, the $\Gamma_9^-$ bands at the $\Gamma$ point continuously shifted down and the band gaps between the $\Gamma_9^-$ and $\Gamma_{6+7}^-$ bands decreased from ~30 to 20, 10 and 0.8 meV, respectively (the sloping blue dashed line in Fig. 3G). Under 1% tensile strain, the $\Gamma_9^-$ bands finally crossed the $\Gamma_{6+7}^-$ bands, forming a type-I Dirac point along the $\Gamma - A$ line (Figs. 3B and 3D, and Fig. S8). The critical tensile strain of the topological phase transition was slightly larger than 0.75% (the vertical blue dashed line in Fig. 3G). The band evolution of the $k_x = 0$ plane from the undoped system (Fig. 3C) to the 1% tensile strain (Fig. 3D) clearly exhibited the emergence of Dirac points along the $k_z$ direction, which was an accidental degeneracy protected by the $C_3$ rotational symmetry (Fig. S8).

In addition to lattice expansion, the external chemical doping also introduced a disorder effect, e.g., changing the electronic structure and Berry curvature by broadening the bands in topological semimetals [39]. Leveraging the experimental lattice constants and the corresponding Mn contents, we investigated the disorder effect of Mn dopant by adopting supercell density functional theory (DFT) calculations for $Mg_{3+\delta}Bi_2:Mn_x$ (Figs. 3E and 3F). As shown in the inset of Fig. 3E, the band crossing occurred in $Mg_{3+\delta}Bi_2:Mn_{0.05}$ at the 0.3% lattice expansion. At a higher doping concentration, i.e., $Mg_{3+\delta}Bi_2:Mn_{0.1}$ with a 0.7% lattice expansion (Fig. 3F), a clear band crossing occurred along the $\Gamma - A$ line (black dashed line in the inset of Fig. 3F). This result suggests that the disorder effect (i.e., band broadening) further accelerated the topological phase transition from a $Z_2$ topological insulator to a Dirac semimetal by decreasing the critical lattice constant. The enlarged purple area of the Dirac semimetal in the phase diagram (Fig. 3G) describes the combined effect of negative chemical pressure and disorder of the Mn dopant, giving rise to low-energy massless excitations around Dirac points with an ultrahigh Fermi velocity and thus a high electron mobility. In contrast, the hole pocket remains almost unchanged during the topological phase transition (Figs. 3E and 3F), which is consistent with our measurements of carrier mobility (Fig. 3H). Overall, we attribute both the giant electron mobility and the enhanced thermoelectric performance in Mn-doped $Mg_{3+\delta}Bi_2$ to the topological phase transition from a $Z_2$ topological insulator to a Dirac semimetal.

**2.4. Enhanced MR and transverse thermopower**

Based on the Boltzmann transport method, we theoretically investigated the MR of Mn-doped $Mg_{3+\delta}Bi_2$ as a function of the external magnetic field: $MR = \frac{\rho(B)-\rho(0)}{\rho(0)} \times 100\%$, where $\rho(B)$ is the electrical resistivity (see Methods in Supporting Information). The Fermi level is set at the Dirac points of the effective band structure $E_f \approx 0.25$ eV, where the carrier mobility is expected to be the highest (Fig. 3F). Fig. 4A and 4B show the MR behaviors of $Mg_{3+\delta}Bi_2$:$Mn_{0.1}$ with an external magnetic field along the *x* and *z* directions, obtained from a $3 \times 3 \times 1$ supercell. We found that the *z*-axis MR exhibited non-saturating behavior when the magnetic field was along the *x*-direction (i.e., B // *x*), reaching 25,000%; in contrast, the *x*-axis and *y*-axis MR were significantly smaller and rapidly saturated with B // *x* (Fig. 4A). When B // *z*, the MR along the *x* and *y* directions degenerated, owing to the in-plane rotational symmetry. In contrast to those of pristine $Mg_3Bi_2$, the *x*-axis and *y*-axis MR of $Mg_{3+\delta}Bi_2$:$Mn_{0.1}$ exhibited non-saturating behavior that reached 3000% (Fig. 4B). Thus, we predicted that the MR of $Mg_{3+\delta}Bi_2$:$Mn_{0.1}$ was one order of magnitude higher than the MR of pristine $Mg_3Bi_2$, which was consistent with our experimental results from the polycrystalline samples (Fig. 2C). The MR is enhanced for two reasons: the increase of mobility (and, hence, longer relaxation time $\tau$) due to the topological phase transition, and the band broadening caused by the disorder effect of the Mn doping.

The transverse thermopower is closely related to the MR, so, without a loss of generality, we considered a case where the temperature gradient was along the *x* direction (i.e., $\nabla_x T \neq 0$, $\nabla_y T =$ $\nabla_z T = 0$) and the magnetic field **B** was along the z direction. The transverse electrical conductivity $\sigma_{xy}$ was generally several orders of magnitude smaller than the longitudinal terms $\sigma_{xx}$ and $\sigma_{yy}$. Thus, the transverse thermopower $S_T$ can be simplified as $S_T \approx \alpha_{yx}/\sigma_{yy} \approx \alpha_{yx} \cdot \rho_{yy}$, where α and ρ are the thermoelectrical conductivity and the electrical resistivity, respectively. Due to the positive relation between transverse thermopower $S_T$ and MR, we naturally expected a huge thermopower enhancement of Mg$_{3+\delta}$Bi$_2$:Mn$_{0.1}$ since the one-order-of-magnitude enhancement of MR was verified. We also examined the general relationship between $S_T$ and MR in other reported materials (Fig. 4C) which further support our conclusion that a huge MR originating from the topological phase transition induces a giant transverse thermopower. Thus, the ongoing search for novel high-performance transverse thermoelectric materials could use the high MR as a quick screen indicator.

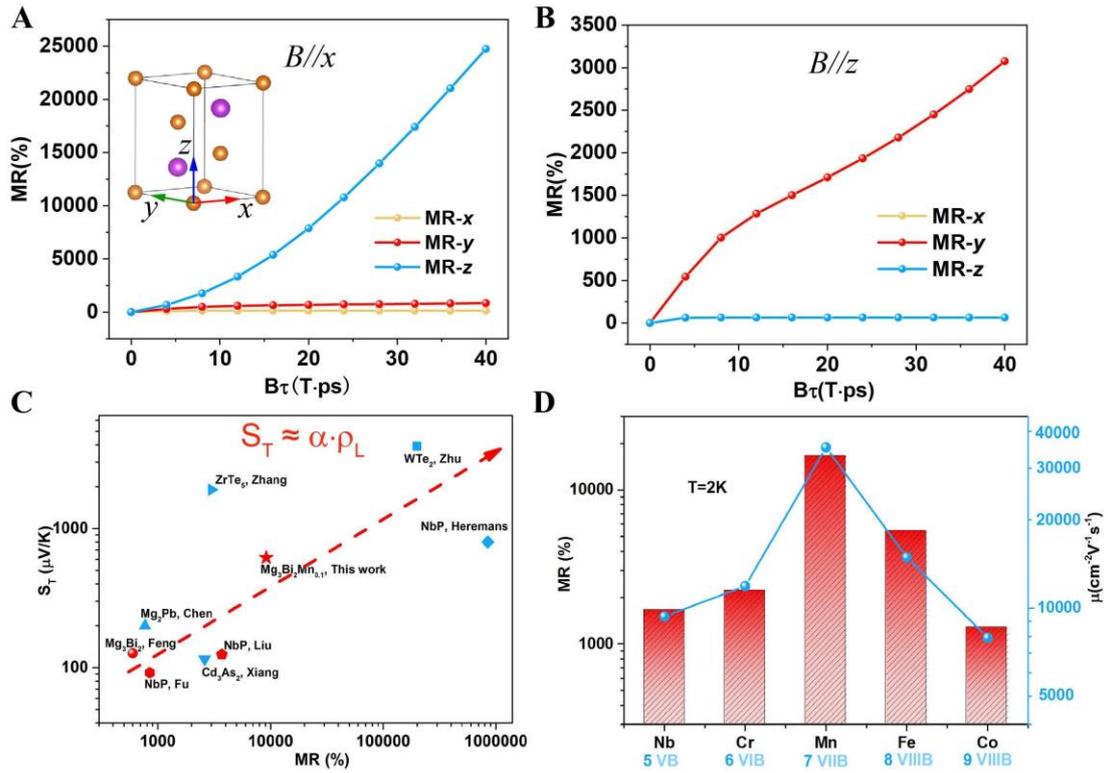

**Fig. 4. Enhanced MR and positive relationship between MR and $S_T$.** MR behaviors of Mg$_{3+\delta}$Bi$_2$:Mn$_{0.1}$ with external magnetic field along the (**A**) *x* and (**B**) *z* directions. The maximum Bτ is set as 40(T·ps), due to a relaxation time larger than that of pristine Mg$_3$Bi$_2$. MR-*x* and MR-*y* are degenerate for B//*z* owing to the in-plane rotational symmetry. (**C**) The generally positive relationship between transverse thermopower $S_T$ and MR for various systems with polycrystalline (red) and single-crystal (blue) materials. (**D**) The MR (left) and carrier mobilities (right) of $Mg_{3+\delta}Bi_2$:$M_{0.1}$ doped with various transition elements (i.e., M = Nb, Cr, Mn, Fe, Co).

To further verify the relationship between the MR and the topological phase transition induced by ionic chemical pressure, we selected four extra transition elements (Nb, Cr, Fe, and Co) with different ionic radii for doping and compared these samples with the $Mg_{3+\delta}Bi_2$:$Mn_{0.1}$. As the doped element evolved from Nb in the 5$^{th}$ group to Co in the 9$^{th}$ group, the MR of $Mg_{3+\delta}Bi_2$:$M_{0.1}$ (M = Nb, Cr, Mn, Fe, Co) started at 1652% (M = Nb), peaked at 16600% (M = Mn), and then decreased to 1274% (M = Co) (Fig. 4D). We measured and fit the Hall resistivity (Fig. S9) to determine the electron mobility of these five transition-elements-doped samples (Fig. 4D). As expected, the electron mobility and MR follow a similar trend, reaching a maximum value for M = Mn. This result is consistent with the situation discussed above in regard to the $Mg_{3+\delta}Bi_2$:$Mn_x$ (x = 0-0.15) samples. Generally, a transition element is introduced in $Mg_{3+\delta}Bi_2$ in the following two cases: (1) to form a four-coordination style at the internal Mg2 position and (2) to form a six-coordination style at the vertex Mg1 and the interstitial positions. The ionic radius for both the four- and six-coordination states of these five different transition elements are shown in Fig. S10 [40]. Notably, the Nb and Cr elements only existed in the form of six-coordination in the crystal. Our results indicate a strong correlation between changes in the ion radius and changes in both electron mobility and MR, regardless of whether the transition elements were distributed in the form of four- or six-coordination states in $Mg_{3+\delta}Bi_2$. These results further confirm the fact that the lattice expansion induces topological phase transition and improves both electron mobility and MR in polycrystalline $Mg_3Bi_2$-based materials.

## 3. Conclusion

In summary, we observed a remarkable transverse thermoelectric effect in a polycrystalline Dirac semimetal Mn-doped $Mg_{3+\delta}Bi_2$. The introduction of a Mn element induced lattice expansion and promoted a topological phase transition towards the Dirac semimetal, rendering the highest MR (16600%) and electron mobility (35280 cm$^2$V$^{-1}$s$^{-1}$) in $Mg_{3+\delta}Bi_2$:$Mn_{0.1}$. Remarkably, the transverse thermopower reached 617 μV/K at 14 K and 14 Tesla, the highest value currently recorded for a polycrystalline material, and the transverse power factor (20393 μWm$^{-1}$K$^{-2}$) exceeds that of pristine $Mg_{3+\delta}Bi_2$ (2182 $\mu Wm^{-1}K^{-2}$) by almost an order of magnitude. Our work offers a general experimental approach to engineering the transverse thermoelectric performance, using dopants to tune the chemical pressure and, in doing so, change the topological electronic structures near the Fermi surface. Our findings also highlight a generally positive relationship between thermopower and MR, which will accelerate the rational design of novel transverse thermoelectric materials.

## 4. Experimental Section/Methods

Experimental details, including materials, characterizations, and DFT calculations, are listed in the Supporting Information.

## Supporting Information

Supporting Information is available from the Wiley Online Library or from the author.

## Acknowledgements

This work was supported by the National Key R&D Program of China (2019YFA0704900), National Natural Science Foundation of China (51872133), Shenzhen Key Program for Long-Term Academic Support Plan (20200925164021002), Guangdong Innovative and Entrepreneurial Research Team Program (2017ZT07C062). The authors would like to thank the support of Core Research Facilities (SCRF) and Center for Computational Science and Engineering of Southern University of Science and Technology, and the support of Guangdong Provincial Key Laboratory Program (2021B1212040001 and 2019B030301001). W.S.L. acknowledges the support from the Tencent Foundation through the XPLORER PRIZE.

## Conflict of Interest:

The authors declare that they have no competing interests.

## Data Availability Statement

The data that support the findings of this study are available from the corresponding author upon reasonable request.